\documentclass[10pt, conference]{IEEEtran}
\IEEEoverridecommandlockouts

\usepackage{cite}
\usepackage{amsmath,amssymb,amsfonts}
\usepackage{algorithmic}
\usepackage{graphicx}
\usepackage{textcomp}
\usepackage[most]{tcolorbox}

\usepackage{multirow}
\usepackage{booktabs}
\usepackage{tabularx}
\usepackage{tablefootnote}
\usepackage{caption}
\usepackage{subcaption}
\usepackage{enumitem}
\usepackage{colortbl}
\usepackage{xcolor}
\usepackage{xspace}
\usepackage{url}
\usepackage[normalem]{ulem}
\usepackage{hyperref}
\usepackage{framed}
\usepackage{subcaption}
\usepackage{listings}
\usepackage{pythonhighlight}
\usepackage{courier}

\usepackage{tikz}
\usepackage{calc}

\usepackage[absolute,overlay]{textpos}

\newcounter{findingctr}

\newcommand{\finding}[2]{\refstepcounter{findingctr}\emph{#1:\label{box:#2}}}

\newlength{\boxw}
\newlength{\boxh}
\newlength{\shadowsize}
\newlength{\boxroundness}
\newlength{\tmpa}
\newsavebox{\shadowblockbox}

\newenvironment{findingenv}[2]%
{\vspace{0.2cm}\noindent
\begin{lrbox}{
\shadowblockbox
}
\begin{minipage}{.98\columnwidth}
\finding{#1}{#2}~}%
{\end{minipage}\end{lrbox}%
\settowidth{\boxw}{\usebox{\shadowblockbox}}%
\settodepth{\tmpa}{\usebox{\shadowblockbox}}%
\settoheight{\boxh}{\usebox{\shadowblockbox}}%
\addtolength{\boxh}{\tmpa}%
\begin{tikzpicture}
\addtolength{\boxw}{\boxroundness * 2}
\addtolength{\boxh}{\boxroundness * 2}

\foreach \x in {0,.05,...,1}
{
\setlength{\tmpa}{\shadowsize * \real{\x}}
\fill[xshift=\shadowsize - 1pt,yshift=-\shadowsize + 
1pt,black,opacity=.04,rounded corners=\boxroundness] 
(\tmpa, \tmpa) rectangle +(\boxw - \tmpa - \tmpa, \boxh - \tmpa - 
\tmpa);
}

\filldraw[fill=white!50, draw=black!80, rounded corners=\boxroundness] (0, 
0) rectangle (\boxw, \boxh);
\draw node[xshift=\boxroundness,yshift=\boxroundness,inner sep=0pt,outer 
sep=0pt,anchor=south west] (0,0) {\usebox{\shadowblockbox}};
\end{tikzpicture}\vspace{0cm}%
}

\definecolor{codegreen}{rgb}{0,0.6,0}
\definecolor{codegray}{rgb}{0.5,0.5,0.5}
\definecolor{codepurple}{rgb}{0.58,0,0.82}
\definecolor{bgcolour}{rgb}{1,1,1}
\definecolor{deleted}{rgb}{0.8, 0.0, 0.0}
\definecolor{added}{rgb}{0.0, 0.9, 0.2}

\lstdefinestyle{pythoncode}{
	language=Python,
	backgroundcolor=\color{bgcolour},   
	commentstyle=\color{codegreen},
	keywordstyle=\color{orange},
	numberstyle=\tiny\color{codegray},
	stringstyle=\color{codepurple},
	basicstyle=\fontfamily{lmss}\scriptsize,
	breakatwhitespace=false,         
	breaklines=true,                 
	captionpos=b,                    
	keepspaces=true,                 
	numbers=left,                    
	numbersep=5pt,
	xleftmargin=0.3cm,                  
	showspaces=false,                
	showstringspaces=false,
	showtabs=false,                  
	tabsize=2,
	frame=top,
	frame=bottom,
	morecomment=[f][\color{deleted}]{-},
    morecomment=[f][\color{added}]{+}
}

\lstdefinestyle{prompt}{
	language=Python,
	backgroundcolor=\color{bgcolour},   
	commentstyle=\color{black},
	keywordstyle=\color{black},
	numberstyle=\tiny\color{black},
	stringstyle=\color{black},
	basicstyle=\fontfamily{lmss}\tiny,
	breakatwhitespace=false,         
	breaklines=true,  
	breakindent=0.5em,               
	captionpos=b,                    
	keepspaces=true,                 
	numbers=left,                    
	numbersep=5pt,
	xleftmargin=0.3cm,                  
	showspaces=false,                
	showstringspaces=false,
	showtabs=false,                  
	tabsize=2,
	frame=top,
	frame=bottom
}

\newcommand{\finalinstances}{\checknum{76}\xspace}

\newcommand{\incorrectinstances}{\checknum{38}\xspace}

\newcommand{\tool}{\textsc{DSChecker}\xspace}
\newcommand{\toolagentic}{\textsc{DSChecker}$_{agent}$\xspace}

\newcommand{\docDirectiveInstances}{\checknum{36}\xspace}

\newcommand{\dirMisuses}{\checknum{18}\xspace}
\newcommand{\dataDepMisuses}{\checknum{20}\xspace}

\newcommand{\llmb}{$P_{base}$\xspace}
\newcommand{\llmdata}{$P_{data}$\xspace}
\newcommand{\llmdir}{$P_{dir}$\xspace}
\newcommand{\llmdataDir}{$P_{full}$\xspace}
\newcommand{\llmfewshot}{$P_{fewshot}$\xspace}

\newcommand{\etal}{et al.}

\newcommand{\so}{Stack Overflow\xspace}
\newcommand{\gh}{GitHub\xspace}
\newcommand{\dsml}{\textcolor{black}{data-centric}\xspace}
\newcommand{\ds}{\textcolor{black}{data science}\xspace}

\newcommand{\esemdataset}{{\small \textsc{DSMisuses}}\xspace}
\newcommand{\dldataset}{{\small \textsc{DLMisuses}}\xspace}
\newcommand{\dl}{DL\xspace}
\newcommand{\datapoint}{code snippet\xspace}
\newcommand{\datapoints}{code snippets\xspace}

\newcommand{\dltool}{{\small \textsc{LLMAPIDet}}\xspace}
\newif\ifshowcomments
\showcommentstrue 

\newcommand\checknum[1]{\ifshowcomments\textcolor{black}{#1}\fi}

\def\BibTeX{{\rm B\kern-.05em{\sc i\kern-.025em b}\kern-.08em
    T\kern-.1667em\lower.7ex\hbox{E}\kern-.125emX}}
\begin{document}

\title{Detecting and Fixing API Misuses of Data Science Libraries Using Large Language Models}

\author{
   \IEEEauthorblockN{Akalanka Galappaththi}
   \IEEEauthorblockA{
   \textit{University of Alberta}\\
   Edmonton, Canada \\
   akalanka@ualberta.ca}
   \and
   \IEEEauthorblockN{Francisco Ribeiro}
   \IEEEauthorblockA{
   \textit{New York University Abu Dhabi}\\
   Abu Dhabi, United Arab Emirates \\
   francisco.ribeiro@nyu.edu}
   \and
   \IEEEauthorblockN{Sarah Nadi}
   \IEEEauthorblockA{
   \textit{New York University Abu Dhabi}\\
   Abu Dhabi, United Arab Emirates \\
   sarah.nadi@nyu.edu}
}

\maketitle

\vspace{-1cm}

\begin{abstract}
Data science libraries, such as scikit-learn and pandas, specialize in processing and manipulating data. 
The data-centric nature of these libraries makes the detection of API misuse in them more challenging. 
This paper introduces \tool, an LLM-based approach designed for detecting and fixing API misuses of data science libraries. 
We identify two key pieces of information, API directives and data information, that may be beneficial for API misuse detection and fixing.
Using three LLMs and misuses from five \ds libraries, we experiment with various prompts. 
We find that incorporating API directives and data-specific details enhances \tool's ability to detect and fix API misuses, with the best-performing model achieving a detection $F_1$-score of 61.18\% and fixing 51.28\% of the misuses.
Building on these results, we implement \toolagentic which includes an adaptive function calling mechanism to access information on demand, simulating a real-world setting where information about the misuse is unknown in advance.
We find that \toolagentic achieves 48.65\% detection $F_1$-score and fixes 39.47\% of the misuses, demonstrating the promise of LLM-based API misuse detection and fixing in real-world scenarios.

\end{abstract}


\section{Introduction}
Most software systems rely on third-party libraries through Application Programming Interfaces (APIs). APIs streamline development, but failing to adhere to API guidelines or assumptions — also known as \textit{API misuse} — can introduce bugs \cite{Detector-Eval-Amann:2019, FUM:2022, API-graph-Zeng:2021}. These misuses can lead to run-time errors, performance issues, or incorrect output \cite{Detector-Eval-Amann:2019, FUM:2022, dl-api-misuse-llm:wei2024, dynamic-misuse-He:2023}. 

Most API misuse research has focused on statically typed languages like Java or C \cite{MuDetect-Amann:2019, Cognicrypt-Kruger:2017, PR-Miner-Li:2005, Jaddet-Wasylkowski:2007}. However, the flexibility of dynamic typing in languages like Python introduces unique challenges for correct API usage~\cite{dynamic-misuse-He:2023}. 
Additionally, the increasing use of Python for data analysis and machine learning introduces new misuse types. 
For example, Wei et al.~\cite{dl-api-misuse-llm:wei2024} identified data conversion errors in Deep Learning (DL) libraries which is a new misuse characteristic.

Building on these findings, Galappaththi et al.~\cite{dc-misuse-smr:24} argued that these characteristics extend beyond \dl libraries and apply to any library with extensive data processing needs, which they refer to as \textit{\dsml libraries}. In essence, libraries with these characteristics are commonly used in \ds applications where extensive data processing, data manipulation, machine learning, or visualization is done. Accordingly, Galappaththi et al.~\cite{dc-misuse-smr:24} study API misuses in five additional non-DL data science libraries (Numpy, pandas, Matplotlib, scikit-learn, and seaborn) and concluded that they suffer from most of the same misuse types found in \dl libraries \cite{dl-api-misuse-llm:wei2024}.

\lstset{
	float=tp,
	style=pythoncode,
	otherkeywords={with}
}
\begin{figure}[t]
    \centering
    \begin{tikzpicture}
        \node[anchor=north west, inner sep=0] (code) at (0,0) {
            \begin{minipage}{\columnwidth}
\begin{lstlisting}[style=pythoncode]
from sklearn.impute import SimpleImputer
import pandas as pd; import numpy as np

df = pd.read_csv("data.csv")

- impt = SimpleImputer(missing_values=np.nan, strategy="mean")
+ impt = SimpleImputer(missing_values=np.nan, strategy="constant", fill_value=0.0001)
imp_array = impt.fit_transform(df)
print(imp_array[:, 1])
\end{lstlisting}
            \end{minipage}
        };
        
        \node[anchor=north east] (tablebox) at (code.north east) {
            \begin{tcolorbox}[colback=gray!10, colframe=black!50, boxrule=0.5pt,
                arc=2pt, left=0pt, right=0pt, top=2pt, bottom=2pt, boxsep=0pt, width=2cm]
                \centering
				\tiny
                \begin{tabular}{l|ll}
                \multirow{5}{*}{\rotatebox{90}{df content}}&\textbf{A} & \textbf{B} \\
                &1 & NaN \\
                &2 & NaN \\
                &$\cdots$ & $\cdots$ \\
                &NaN & NaN \\
                \end{tabular}
            \end{tcolorbox}
        };


    \end{tikzpicture}
    \vspace{-0.6cm}
    \caption{\texttt{SimpleImputer} misuse~\cite{dc-misuse-smr:24}: \texttt{strategy="mean"} drops column \texttt{B} as it contains only \texttt{NaN} values, causing an indexing error when accessed (from Stack Overflow \href{https://stackoverflow.com/questions/60527883/does-simpleimputer-remove-features}{60527883}).}
    \label{fig:imputer}
    \vspace{-0.6cm}
\end{figure}

Figure~\ref{fig:imputer} shows a misuse from scikit-learn in a diff format, highlighting the original code containing the API misuse and the corrected version~\cite{dc-misuse-smr:24}. 
The behavior of \texttt{SimpleImputer}~\cite{sk-simpleimputer} depends on both the input data and the selected imputation strategy, as constrained by the following \textit{API directive}~\cite{Directives-Monperrus:2012}: \textit{``Columns which only contained missing values at \texttt{fit} are discarded upon \texttt{transform} if strategy is not `constant'.''}~\cite{sk-simpleimputer}.
Accordingly, given the original API call on Line 6, the missing values in the first column of \texttt{imp\_array} get replaced by its mean; however, the second column which contains only missing values (i.e., \texttt{NaN}) will be discarded. 
Subsequently, the print statement on Line 9 raises an \texttt{IndexError} because \texttt{imp\_array} has only one column, making access to index 1 invalid. 
One potential fix is to change the \texttt{strategy} parameter to \texttt{"constant"} and introduce a \texttt{fill\_value} parameter with a constant argument, as shown on Line 7.
Galappaththi \etal~\cite{dc-misuse-smr:24} defined this type of misuses as \textit{data-dependent misuses}, because the original code works if column \texttt{B} contained at least one numerical value.

Previous research showed that API misuses are a prevalent problem and the majority of them cause program crashes \cite{Misuse-Monperrus:2013, MuDetect-Amann:2019}.
Additionally, detecting API misuses in libraries that involve a lot of data processing is challenging, because we have to account for the internal composition of complex data structures (e.g., DataFrames or arrays) that these APIs process.
With the known limitations of static analysis in capturing such dynamic information~\cite{dl-api-misuse-llm:wei2024} and the known shortcomings of traditional pattern-based API misuse detectors~\cite{GrouMiner-Nguyen:2009,  Jaddet-Wasylkowski:2007, Colibri-ML-Lindig:2015, PR-Miner-Li:2005, Misuse-Monperrus:2013}, we need new ways for detecting misuses of \ds APIs. 
With recent developments in large language models (LLMs), Wei \etal~\cite{dl-api-misuse-llm:wei2024} experiment with LLMs to detect and fix \dl API misuses. 
Motivated by their work, in this paper, we investigate whether LLMs can detect and fix API misuses in \ds libraries. 
Our key expectation is that determining the correct API usage in this context would likely require knowledge of the API directives as well as dynamic information about the data being processed. 
Accordingly, we design an LLM-based API misuse detection and fixing approach, \tool, that employs a simple prompting strategy, leveraging both static information retrieved from the API documentation and dynamic information obtained by running the code.
To evaluate our approach, we conduct an empirical study on five \ds libraries, answering the following questions:

\newcommand{\rqonetext}{Which information is most effective in guiding LLMs to detect and fix API misuses of \ds libraries?}
\newcommand{\rqtwotext}{Does few-shot prompting improve LLMs' API misuse detection and fixing?}
\newcommand{\rqthreetext}{Can \tool detect/fix misuses of other \ds libraries?}
\newcommand{\rqfourtext}{How does \tool compare to other LLM-based API misuse detection/fixing tools?}
\newcommand{\rqfivetext}{Can \toolagentic detect and fix API misuses?}

\begin{itemize}[leftmargin=*,align=left]
    \setlength{\parindent}{0pt} 
    \item[\textbf{RQ1}:] \textbf{\rqonetext} We conduct an ablation study using zero-shot prompts to assess the role of API directives and data context.

    \item[\textbf{RQ2}:] \textbf{\rqtwotext} After finding the best zero-shot prompt, we test if few-shot prompts lead to improvement.

    \item[\textbf{RQ3}:] \textbf{\rqthreetext} We go beyond the five initial libraries by using 15 \dl misuses from Wei et al.~\cite{dl-api-misuse-llm:wei2024}'s dataset.

    \item[\textbf{RQ4}:] \textbf{\rqfourtext} We compare \tool against \dltool~\cite{dl-api-misuse-llm:wei2024}, an existing LLM-based method for detecting and fixing \dl API misuses.

    \item[\textbf{RQ5}:] \textbf{\rqfivetext} We explore if LLMs equipped with function calling (\toolagentic) can autonomously retrieve API documentation and data information in real-world scenarios, where this is not readily available.
\end{itemize}

Our results show that incorporating API directives and variable information significantly improves LLMs' performance with an average 7\% increase in $F_1$-score for detecting and a 5\% increase in fix rate for fixing \ds API misuses.  \tool also outperformed a prior misuse detection tool for \dl libraries~\cite{dl-api-misuse-llm:wei2024} (46.15\% vs. 16.49\%). Additionally, while \toolagentic's performance drops in comparison to \tool, it shows potential of real-world application where precise information may not be readily available.

In summary, our contributions include:
\begin{enumerate}[leftmargin=*,align=left]
\item A new LLM-based approach, \tool, that combines API directives and dynamic data information for enhanced API misuse detection and fixing.
\item Evaluation of our approach on an existing \ds API misuse dataset \cite{dc-misuse-smr:24}, including an ablation study to understand the effect of the different prompt information and experimenting with prompting techniques.
\item A comparison of \tool to a different LLM-based misuse detection/fixing approach previously developed for DL libraries as well as the evaluation of \tool on additional DL misuses.
\item Implementation and evaluation of an agentic version of \tool, \toolagentic, for the practical application of LLM-based detection and fixing where no prior information about the misuse is known.
\end{enumerate}

Our replication package contains our results as well as the source code and data to run all our experiments: \href{https://doi.org/10.6084/m9.figshare.28327148}{https://doi.org/10.6084/m9.figshare.28327148}. 

\section{Background and Related Work}
\label{sec:related-work}
\vspace{-0.2cm}

\paragraph{API Misuse}
An \textit{API misuse} occurs when there is any deviation from the API’s intended usage, such as failing to invoke a required method~\cite{dynamic-misuse-He:2023, dl-api-misuse-llm:wei2024, FUM:2022, Detector-Eval-Amann:2019}, using unsupported parameter values~\cite{dynamic-misuse-He:2023}, or passing incorrect data types~\cite{dynamic-misuse-He:2023}.
API misuses can cause run-time errors, performance issues, or incorrect outputs \cite{dynamic-misuse-He:2023, dl-api-misuse-llm:wei2024, FUM:2022, Detector-Eval-Amann:2019}.
In this paper, we focus on \ds libraries that process, analyze, and derive insights from data. These libraries typically accept diverse data structures such as pandas DataFrames or NumPy arrays and have complex processing workflows~\cite{dc-misuse-smr:24}.

\paragraph{API Misuse Detection}
Most of the early work on detecting API misuses is based on mining frequent usage patterns from source code to identify correct usages with deviations from those mined patterns considered as misuses \cite{Misuse-Monperrus:2013, PR-Miner-Li:2005, Colibri-ML-Lindig:2015, GrouMiner-Nguyen:2009, Jaddet-Wasylkowski:2007, MuDetect-Amann:2019}. 
However, these approaches typically suffer from low precision and recall~\cite{Detector-Eval-Amann:2019}. 
To address this, researchers proposed automatically learning correct usages from API documentation \cite{API-KG-Ren:2020, API-graph-Zeng:2021}. However, this typically relies on static parameters and return types to define usage constraints, making it effective for statically typed languages like Java but not suitable for Python.
For example, Matplotlib's \texttt{plot} accepts various parameter types for \texttt{x} and \texttt{y}.
Accordingly, this creates runtime-dependent constraints for subsequent calls to additional APIs such as \texttt{axvspan}.
To the best of our knowledge, static fine-grained API usage graphs~\cite{API-KG-Ren:2020, API-graph-Zeng:2021} cannot capture such downstream dynamic API typing dependencies, making them inapplicable to our work. 
Another misuse detection direction uses domain specific languages that encode usage constraints, which are either automatically extracted from the documentation or manually specified by an expert \cite{Cognicrypt-Kruger:2017, IM-Gu:2019, SSL-Li:2019, gulami-thesis:2022}. 

Given the above limitations, researchers started leveraging LLMs for API misuse detection.
For example, Baek et al.~\cite{llm-crypto:24} demonstrated the effectiveness of LLMs for detecting Java cryptographic API misuses, showing that their fine-tuned model outperformed state-of-the-art rule-based detectors in this domain \cite{crypto-android:18, crypto-android:13, crysl:21}. In contrast, our work uses additional contextual information for off-the-shelf LLMs, focusing on a different domain (\ds) and language (Python). 

Wei et al.\cite{dl-api-misuse-llm:wei2024} study API misuses in PyTorch and TensorFlow. They construct a dataset of 891 misuses from version-control history (\dldataset), and categorize them by type (e.g., missing method), cause (e.g., device mismanagement), and symptom (e.g., runtime error).
They propose \dltool, an LLM-based detection and fixing approach with three steps:
(1) Few-shot prompting is used to extract fixing rules from 600 before-and-after misuse examples; the remaining 291 are reserved for evaluation;
(2) To detect misuses, the LLM generates a natural language summary of a code snippet, which is compared to existing rules to retrieve the top four matches; a match constitutes a misuse detection;
(3) If a misuse is found, the LLM is prompted with the code and retrieved rules to produce an explanation and a fix.
While \dltool outperforms a rule-based detector\cite{TADAF:2022}, it still achieves low recall (16.49\%) and has a 23\% fix success rate.
Our approach uses API documentation and dynamic variable information instead of rule mining, simplifying the pipeline.
As DL libraries fall under \dsml, we evaluate generalizability in RQ3 (applying \tool to \dldataset) and relative performance in RQ4 (applying both tools to a shared subset).

\paragraph{Code generation and program repair with LLMs}
LLMs have demonstrated effectiveness in understanding code and generating natural language explanations for code snippets \cite{llm-code-understanding:24}. Leveraging these capabilities, LLMs have been applied to generate code snippets for detecting bugs in DL applications \cite{llm-zero-shot-fuzzer:23}. Guan et al.~\cite{llm-context-learning:24} used contextual information to improve LLM's performance in generating test cases to detect model optimization bugs in DL applications.
Similarly, we also provide LLMs with contextual information (e.g., API directives) but for a different problem and beyond only DL applications.

In the context of automatic program repair, Xia et al.~\cite{llm-apr-diff-patch-setting:23} evaluated different prompt settings for generating fixes.  
They find that generating fixes for buggy lines is more effective than reconstructing the entire function. We also generate fixes to replace buggy lines using a unified diff format. However, unlike Xia et al.~\cite{llm-apr-diff-patch-setting:23}, we do not provide prefix or suffix indicators to highlight bug locations, as part of our goal is to assess the LLMs' ability to identify problematic usage.

\paragraph{LLM agents}
\textit{LLM agents} are autonomous applications interacting with external tools without direct human intervention \cite{agents:25}. \textit{Function calling} is a mechanism that empowers these agents to communicate with code or external services, enabling them to retrieve information or perform specific tasks \cite{tool-calling}. 
For example, \textit{WebGPT}~\cite{webGPT} enabled web browsing for generating comprehensive answers, while \textit{ReAct}\cite{react:23} explored reasoning and action-taking, such as planning and interacting with external resources. While we do not focus on complex reasoning, we leverage a form of agency by allowing LLMs to call functions, enabling direct interaction with code.
\vspace{-0.2cm}
\section{\tool: An LLM-based Misuse Detection and Fixing Approach}
\label{sec:approach}
 
\begin{figure}[t!]
	\vspace{-0.4cm}
	\includegraphics[width=0.5\textwidth]{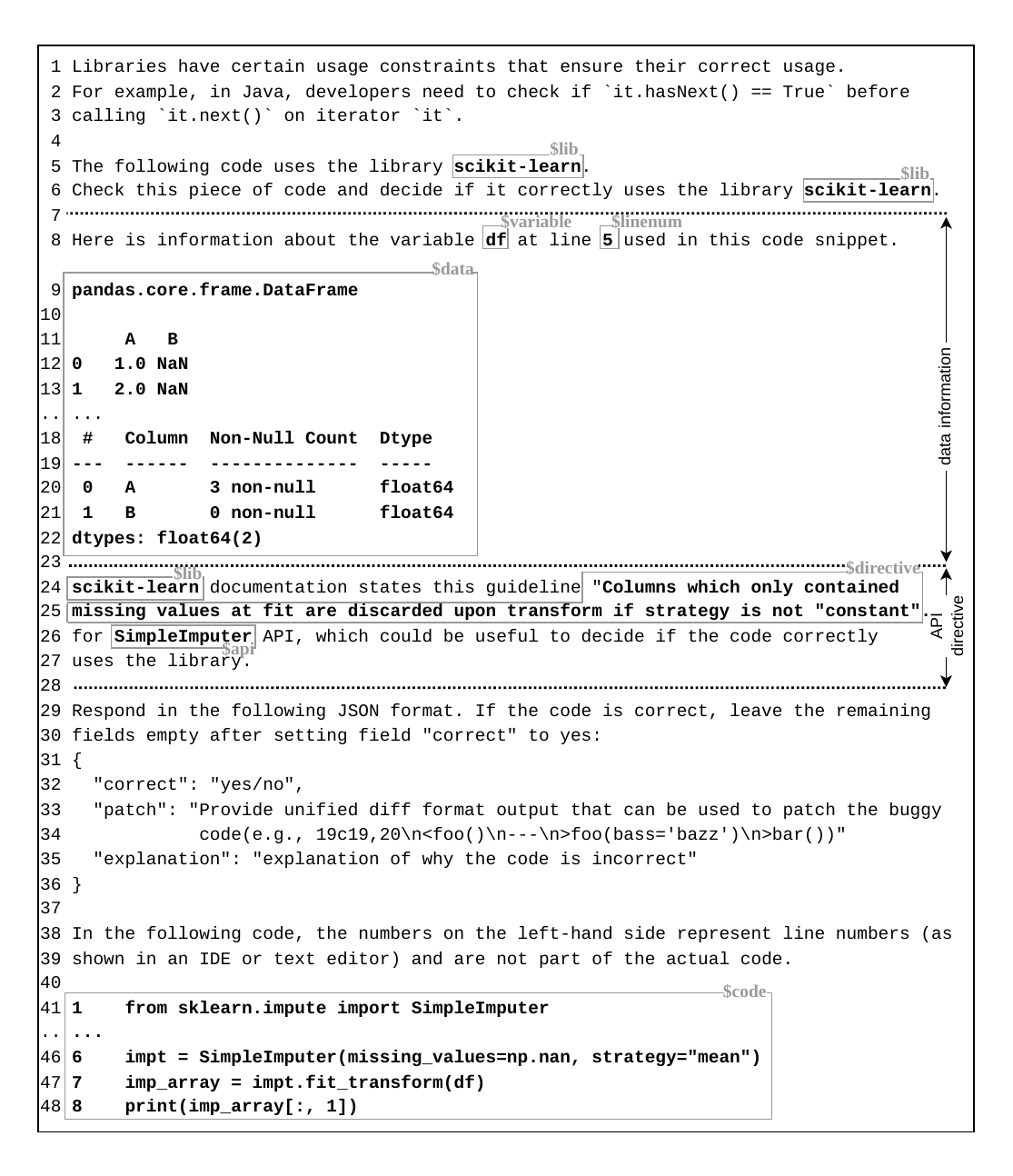}
	\vspace{-0.7cm}
	\caption{Example of prompt provided to an LLM to detect and fix the misuse in Figure~\ref{fig:imputer}}
	\label{fig:full-prompt-simpleimputer}
	\vspace{-0.6cm}
\end{figure}

\tool is based on the simple idea of prompting an LLM with the right information.
It takes a piece of code that uses third-party \ds APIs that we want to check.
The output indicates whether there is a misuse in the code and if so, how to fix it.
Figure~\ref{fig:full-prompt-simpleimputer} shows an example of the prompt \tool uses to detect and fix the misuse in Figure~\ref{fig:imputer}.
LLMs have demonstrated natural language understanding capabilities \cite{llm-nl-understanding:pci23} as well as code understanding capabilities \cite{llm-code-understanding-myers:icse24}.
Thus, our first key insight is that if an API has a directive describing its expected usage, an LLM can potentially check if the directive is being followed in a given piece of code.
In Figure~\ref{fig:full-prompt-simpleimputer}, lines 24-27 show this API directive in the prompt. 
The second insight is that since these APIs process a lot of data and the nature of this input data can influence whether the API behaves correctly~\cite{dc-misuse-smr:24}, providing the LLM with information about the passed data, may help it determine the correct API usage.
Lines 8-22 shows the data information for this example.

The prompt asks the model to use a JSON object as the response format. If the LLM detects a misuse, the output will follow the format: \texttt{\{"correct": "no", "patch": "\dots", "explanation": "\dots"\}}. 
Also, we instruct the LLM to generate a patch in unified diff format, and a free text explanation. If no misuse is found, the model simply responds with \texttt{\{"correct": "yes"\}}.

\vspace{-0.2cm}
\section{Dataset Preparation and Evaluation Metrics}
\label{sec:dataset}

\subsection{Dataset Preparation: \esemdataset}
\label{sec:data-prep}
\vspace{-0.2cm}

We use the API misuse dataset by Galappaththi et al.~\cite{dc-misuse-smr:24}.
We now describe the original dataset and our extensions of it.

\paragraph*{Original Data} Galappaththi et al.~\cite{dc-misuse-smr:24}'s dataset contains 49 API misuses collected from Stack Overflow posts and GitHub commits related to five \ds libraries: NumPy, pandas, Matplotlib, scikit-learn, and seaborn.
Each misuse comes with a minimal reproducible example to illustrate the misuse, and indication of whether it is data dependent.
If the misused API has an explicit directive, typically found in API references or occasionally in user guides and examples from the online library documentation, it is also included. 

\paragraph*{Vet and Reproduce} We first carefully review the provided data to make sure we can verify the misuse and reproduce the error or incorrect behavior through the provided code example.
Based on this review, we remove 8 misuses that we could not reproduce, leaving us with 41 misuses.
We also remove 3 duplicate misuses (e.g., two instances of seaborn's \texttt{set\_palette()} API misuse extracted from two different Stack Overflow posts). This leaves us with \incorrectinstances misuses, out of which \dirMisuses are misuses of an API with an explicit usage directive while \dataDepMisuses are data dependent.
  
\paragraph*{Replace Hard-coded Data} 
We replace hard-coded data, such as DataFrames created from dictionaries, with data loaded from a file to assess the effect of providing data information.

\paragraph*{Add data information} \tool prompts require information about the data processed by the API, which the original dataset does not provide. To extract this information, we manually inspect the code to identify relevant variables, and then instrument it with \texttt{print(type(var))} statements to identify variable types. We find three main data structures that appear in the \datapoints: DataFrames, NumPy arrays, and lists. For DataFrames, we additionally use \texttt{print(var.info())} and \texttt{print(var.head(3))} to record column names, types, and sample rows respectively. For NumPy arrays, we use \texttt{print(var.shape, var.dtype)} to capture the shape and data type of the array. For lists, we use \texttt{print(len(var))} to record list length. Running the instrumented code yields outputs such as type, and associated content. For instance, for the variable \texttt{df} at line \texttt{4} in Figure~\ref{fig:imputer}, the output includes its type DataFrame, sample values, and column metadata.

\paragraph*{Create correct usages} 
To evaluate effectiveness in identifying both correct and incorrect API usages, we augment the dataset with a fixed code snippet for each of the \incorrectinstances misuses.
We refer to the final updated dataset as \esemdataset which contains \finalinstances code snippets.
\vspace{-0.2cm}

\subsection{Evaluation Metrics}
\label{sec:metrics}
\vspace{-0.2cm}

In all research questions, our evaluation focuses on two primary aspects: misuse detection and fixing. 
Since the metrics are common across all RQs, we discuss them here first.

\subsubsection{Misuse detection metrics}

When evaluating \tool's ability to detect API misuses, we consider that it correctly detects a misuse if it flags the code snippet as not correct (i.e., outputs \texttt{"correct":"no"}) \textit{and} provides an accurate explanation about the misuse.  
We first automatically compare the \texttt{correct} field of the model's response to the ground truth in \esemdataset.
We then manually review the explanations of misuses correctly marked as \texttt{no}.
If the explanation accurately describes the misuse, we confirm this as a correct detection; otherwise, we mark it as not detected.
We follow this two-step process as we found that relying solely on the yes/no classification is insufficient; LLMs occasionally flag a code snippet as having a misuse but then describe non-existent issues or do not provide adequate explanations for the detected problems. 

At the end of this process, we calculate precision ($P$), recall ($R$), and $F_1$-score for misuse detection.
We calculate \textit{precision} as the number of correctly detected misuses divided by the total number of code snippets flagged as misuse by the tool and \textit{recall} as the total number of correctly detected misuses divided by the total number of misuses in the dataset. 
We calculate \textit{$F_1$-score} as the harmonic mean of the precision and recall: $F_1 = 2*P*R/(P+R)$.

\subsubsection{Misuse fixing metrics}
To evaluate fixing, we automatically apply the LLM-provided patch to the corresponding code snippet and execute the patched code. 
We then manually evaluate whether the patched code resolves the original error or produces the correct output. 
For misuses that result in errors, we consider a fix to be correct \textit{only} if it resolves the original error without introducing additional errors.
For misuses that result in incorrect output, we ensure that the problem has been fixed based on the misuse description in the dataset.
We calculate \textit{Fix Rate} as the proportion of correct patches relative to the total number of misuses (\incorrectinstances for \esemdataset). 

\subsubsection{Bootstrap sampling}
To evaluate whether the observed performance differences between LLMs and prompt variations are statistically significant, we apply bootstrap sampling~\cite{bootstrap:11}. For each setting, we repeatedly sample 20 code snippets 50 times and compute $F_1$-scores and fix rates for each sample, yielding performance distributions. After confirming non-normality with the Shapiro–Wilk test~\cite{shapiro:11}, we apply Dunn’s test with Bonferroni adjustments~\cite{dunn:2009} at a 95\% confidence level ($\alpha = 0.05$) to identify statistically significant differences between distribution pairs.

\vspace{-0.2cm}
\section{RQ1: Which information is most effective in guiding LLMs to detect and fix API misuse?}
\label{sec:zero-shot}

To answer RQ1, we conduct an ablation study to compare the effect of the different prompt components from Figure \ref{fig:full-prompt-simpleimputer}.

\vspace{-0.2cm}
\subsection{RQ1 Setup}
\label{sec:rq1-setup}
\vspace{-0.2cm}

Figure~\ref{fig:full-prompt-simpleimputer} shows the full prompt with an API directive and data information. 
We generated this prompt using a prompt template that contains fixed text and placeholders that we substitute values for according to the target code snippet. 
Figure~\ref{fig:full-prompt-simpleimputer} shows these placeholders in gray rectangles (e.g., \texttt{\$lib} is a placeholder for library name), while showing the substituted text for the example in bold.
For easier visualization, the dotted lines separate the sections that contain data information (Lines 8-22) versus the API directive (Lines 24-27).
We now discuss the four variations of the prompt used in RQ1.
Note that these are all zero-shot prompts where no examples are provided.

\subsubsection{Prompt with API directives and data information (\llmdataDir)}
This is the full prompt template with all available information in the sections labeled ``data information'' and ``API directive''. To generate a concrete prompt for a given code snippet, we replace the placeholders \texttt{\$lib} and \texttt{\$code} with the corresponding library name and code snippet, respectively.
The ``data information'' section has three placeholders: \texttt{\$variable} and \texttt{\$linenum}, which are replaced with the corresponding variable name and line number respectively, and \texttt{\$data}, which is replaced with the type of the variable, sample values, and any additional information about the variable, as recorded in the dataset (See Section~\ref{sec:data-prep}).

In Figure~\ref{fig:full-prompt-simpleimputer}, ``API directive'' section has three placeholders: \texttt{\$lib}, \texttt{\$directive}, and \texttt{\$api}. We replace them with the library name, the API directive, and the corresponding API name, respectively.
The presentation of this section varies according to the directive's context. For example, if a directive is specific to a parameter, we mention the parameter name.

Note that a code snippet may contain multiple variables and APIs, but in RQ1 (and later RQ2), we focus only on those relevant to the potentially misused API to isolate the effect of providing the LLM with different information. 

\subsubsection{Prompt with only API directives (\llmdir)}
This variant disregards data information and provides only API directives.

\subsubsection{Prompt with only data information (\llmdata)}
This variant ignores API directives and provides only data information.

\subsubsection{Base prompt (\llmb)}
The base prompt is the control prompt of RQ1 which establishes a baseline for evaluating the effect of additional information on detecting and fixing misuses.
It strips out all additional information on Lines 8-27 and provides only the code snippet and the library being used. 

\paragraph*{Selected LLMs}
We select three LLMs for our experiments based on their performance in recent studies~\cite{llm-static-bug-detect:24, llm-context-learning:24}: two proprietary models--OpenAI’s gpt-4o-mini-2024-07-18 and gpt-4o-2024-05-13--and one comparably sized open-source model, Meta’s llama-3.1-405b-Instruct. 
As a short-hand, we refer to these models as \texttt{4o-mini}, \texttt{4o}, and \texttt{llama} respectively.

\vspace{-0.2cm}
\subsection{RQ1 Results}
\label{sec:rq1-results}

\begin{table}
    \centering
        \caption{\tool's misuse detection (RQ1-2)}
        \vspace{-0.2cm}
        \resizebox{0.95\columnwidth}{!}{
        \begin{tabular}{llrrrr|r}
            \toprule
            \multirow{2}{*}{\textbf{Metric}} &
            \multirow{2}{*}{\textbf{Model}} &
            \multicolumn{4}{c}{\textbf{Zero-shot (RQ1)}} &
            \multicolumn{1}{c}{\textbf{Few-shot (RQ2)}} \\
            \cmidrule(lr){3-6}\cmidrule(lr){7-7}
            & & \boldmath{\llmb} & \boldmath{\llmdata} & \boldmath{\llmdir} & \boldmath{\llmdataDir} & \boldmath{\llmfewshot} \\
            \midrule
            
            	\rowcolor{gray!20}
            \multicolumn{7}{c}{All data (\finalinstances)} \\
            
            \midrule
                \multirow{3}{*}{P} &
                4o-mini   & \textbf{45.65\%} & 37.04\% & 38.46\% & 33.33\% & {41.67\%} \\
                &4o     & 48.78\% & 51.16\% & 52.38\% & \textbf{55.00\%} & \textbf{57.14\%} \\
                & llama & 48.89\% & 50.00\% & 52.17\% & \textbf{56.52\%} & \textbf{56.82\%} \\
                \midrule
                \multirow{3}{*}{R} &
                4o-mini   & \textbf{53.85\%} & 51.28\% & 51.28\% & 46.15\% & \textbf{64.10\%} \\
                & 4o     & 51.28\% & \textbf{56.41\%} & \textbf{56.41\%} & \textbf{56.41\%} & \textbf{61.54\%} \\
                & llama & 56.41\% & 48.72\% & 61.54\% & \textbf{66.67\%} & 64.10\% \\
                \midrule
                \multirow{3}{*}{F1} &
                4o-mini   & \textbf{49.41\%} & 43.01\% & 43.96\% & 38.71\% & \textbf{50.51\%} \\
                & 4o     & 50.00\% & 53.66\% & 54.32\% & \textbf{55.70\%} & \textbf{59.26\%} \\
                & llama & 52.38\% & 49.35\% & 56.47\% & \textbf{61.18\%} & 60.24\% \\
            \midrule
            
            	\rowcolor{gray!20}
            \multicolumn{7}{c}{{Potentially misused API has a directive (\docDirectiveInstances)}}\\

			\midrule
                \multirow{3}{*}{P} &
                4o-mini   & \textbf{56.52\%} & 48.28\% & 41.94\% & 40.00\% & {45.45\%} \\
                & 4o     & 54.17\% & 54.55\% & 59.09\% & \textbf{65.00\%} & \textbf{66.67\%} \\
                & llama & 50.00\% & 50.00\% & 56.52\% & \textbf{60.00\%} & 59.09\% \\
                \midrule
                \multirow{3}{*}{R} &
                4o-mini   & 68.42\% & \textbf{73.68\%} & 68.42\% & 63.16\% & \textbf{78.95\%} \\
                & 4o     & \textbf{68.42\%} & 63.16\% & \textbf{68.42\%} & \textbf{68.42\%} & \textbf{73.68\%} \\
                & llama & 57.89\% & 42.11\% & 68.42\% & \textbf{78.95\%} & 68.42\% \\
                \midrule
                \multirow{3}{*}{F1} &
                4o-mini   & \textbf{61.90\%} & 58.33\% & 52.00\% & 48.98\% & {57.69\%} \\
                & 4o     & 60.47\% & 58.54\% & 63.41\% & \textbf{66.67\%} & \textbf{70.00\%} \\
                & llama & 53.66\% & 45.71\% & 61.90\% & \textbf{68.18\%} & 63.41\% \\
            \bottomrule

        \end{tabular}
        }
    \label{tab:detect-eval}
    \vspace{-0.5cm}
\end{table}

\vspace{-0.2cm}
\subsubsection{Misuse detection}
Table~\ref{tab:detect-eval} shows RQ1 detection results.
\paragraph{Full dataset} 
We first focus on the top left part of Table~\ref{tab:detect-eval}, which presents the zero-shot results over all \finalinstances~\datapoints.
Looking at the $F_1$-score, we find that for both \texttt{4o} and \texttt{llama}, \llmdataDir is most effective at detecting misuses. However, \texttt{4o-mini} performed best with \llmb and had its worst performance when provided \llmdataDir.

We find that while \texttt{4o-mini} correctly stated the presence of misuses for \checknum{35} misuses, it failed to provide adequate explanations for most of them when provided any additional information, resulting in a lower number of true positives and thus lower recall and precision.
For example, for the misuse in Figure \ref{fig:imputer}, \texttt{4o-mini} with \llmdataDir{} claimed that passing the entire DataFrame is not allowed, despite \texttt{SimpleImputer} accepting pandas DataFrames.
On the other hand, \texttt{4o} and \texttt{llama} flagged this misuse and provided the right explanation.
These findings confirm that some models are more effective than others in interpreting provided information~\cite{xu2024retrieval, liu2023lostmiddlelanguagemodels}.

We find that the performances of \texttt{4o} and \texttt{llama} is significantly higher than \texttt{4o-mini} (\textit{p}\textless{}0.05). However, the performance between \texttt{4o} and \texttt{llama} is not significant (\textit{p}=0.468).

\paragraph{Code snippets with API directives} 
If a \datapoint does not have an API directive to include in the prompts, \llmdir and \llmdataDir will be equivalent to \llmb and \llmdata, respectively.
This may blur our understanding of the effect of API directives.
Accordingly, we next focus \textit{only} on the subset of \docDirectiveInstances \datapoints in \esemdataset where the potentially misused API has an explicit directive such that the generated prompts are different.
This is shown in the bottom left part of Table \ref{tab:detect-eval}. 
When compared to \llmb, \llmdir increased (or kept) the recall of all models and the precision and $F_1$-score of two of the three models.
For both \texttt{4o} and \texttt{llama}, adding directives significantly improves their ability to detect API misuses (\textit{p}\textless{}0.05). 
These results further highlights the positive effect of providing the API directive to the LLM.

\begin{figure}[t!]
    \centering
    \includegraphics[width=0.49\textwidth]{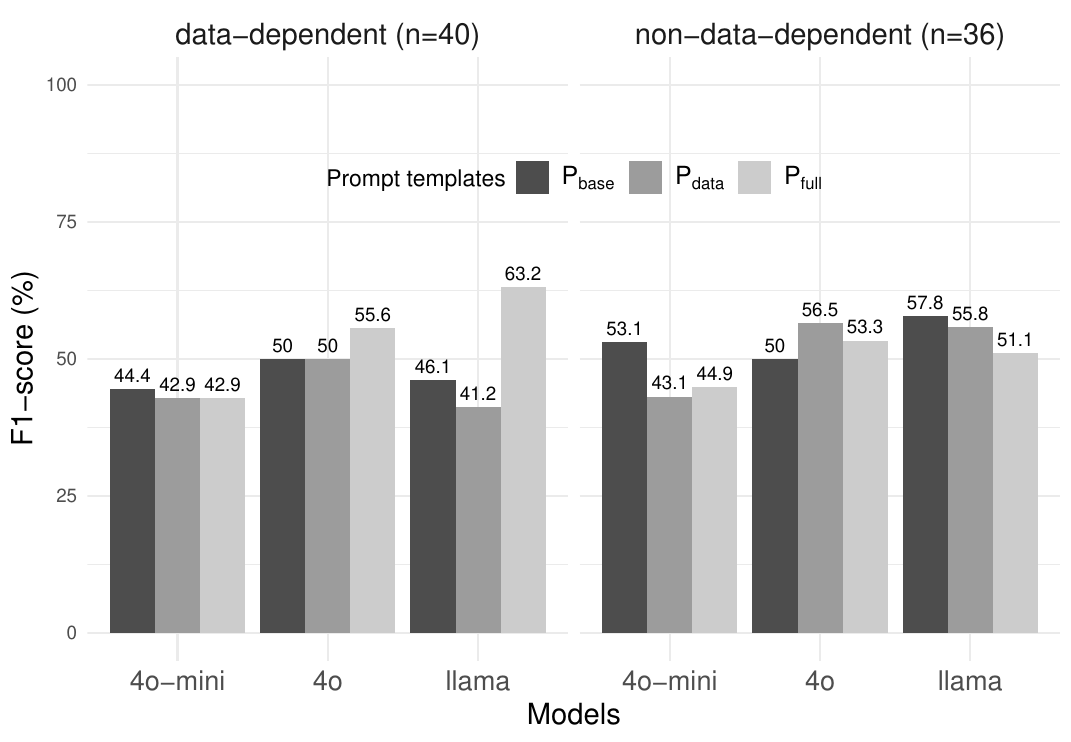}
    \vspace{-0.5cm}
    \caption{Effect of adding data to data-dependent misuses vs. non-data-dependent misuses.}
    \label{fig:data-effect}
    \vspace{-0.6cm}
\end{figure}

\paragraph{Data-dependent misuses} 
Recall that not all misuses are data dependent.
It could be the case that providing data information helps only when the potential misuse is data dependent but confuses the model otherwise.
Thus, we further compare the performance on subsets of data-dependent and non-data-dependent \datapoints.

Figure \ref{fig:data-effect} presents the $F_1$-scores for the three models across data-dependent (40) and non-data-dependent (36) \datapoints using the prompt templates \llmb, \llmdata, and \llmdataDir. 
Given the negative impact of adding information on \texttt{4o-mini}, we focus our analysis on \texttt{4o} and \texttt{llama}, for which Table \ref{tab:detect-eval} indicates \llmdataDir is generally most effective.
In the data-dependent facet, adding only data information (\llmdata) did not improve, or even decreased, $F_1$-scores for these two models.
However, \llmdataDir always increased the $F_1$-score on these data-dependent \datapoints, with a particularly large increase for llama. This increase is statistically significant for both \texttt{4o} and \texttt{llama} (\textit{p}\textless{}0.05).
On the other hand, for non-data-dependent code snippets, we find that \llmdataDir resulted in lower $F_1$-scores for llama compared to \llmb, and for \texttt{4o} compared to \llmdata.
This does suggest that the data and directive combination sometimes hurts when the API usage is not data dependent. 
However, since the nature of the API usage is unknown beforehand and the benefit on data-dependent misuses is much higher than the slight performance loss on the other \datapoints, we conclude that \llmdataDir is still worthwhile.

\begin{table}[!t]
    \centering
        \caption{\tool's fix rate (RQ1-2)}
        \vspace{-0.2cm}
        \resizebox{0.80\columnwidth}{!}{
            \begin{tabular}{lrrrr|r}
                \toprule
                \multirow{2}{*}{\textbf{Model}} & 
                \multicolumn{4}{c}{\textbf{Zero-shot (RQ1)}} & 
                \multicolumn{1}{c}{\textbf{Few-shot (RQ2)}} \\
                \cmidrule(lr){2-5}\cmidrule(lr){6-6}
                & 
                \boldmath{\llmb} & 
                \boldmath{\llmdata} & 
                \boldmath{\llmdir} & 
                \boldmath{\llmdataDir} & 
                \boldmath{\llmfewshot}  
                \\
                \midrule
                	
                	\rowcolor{gray!20}
                \multicolumn{6}{c}{All misuses (\incorrectinstances)} \\
                \midrule
                4o-mini & \textbf{43.59\%} & 38.46\% & 38.46\% & 38.46\% & {41.03\%} \\
                4o & 41.03\% & 35.90\% & \textbf{46.15\%} & 43.59\% & \textbf{53.85\%} \\
                llama & 46.15\% & 35.90\% & 46.15\% & \textbf{51.28\%} & \textbf{51.28\%} \\

                \midrule
                \rowcolor{gray!20}
                \multicolumn{6}{c}{Misuses where misused API has a directive (\dirMisuses)} \\
                \midrule
                4o-mini & \textbf{52.63\%} & \textbf{52.63\%} & \textbf{52.63\%} & 47.37\% & \textbf{52.63\%} \\
                4o & 47.37\% & 42.11\% & 57.89\% & \textbf{63.16\%} & \textbf{63.16\%} \\
                llama & 47.37\% & 31.58\% & 52.63\% & \textbf{57.89\%} & 52.63\% \\

                \bottomrule
            \end{tabular}
        }    
    \label{tab:fix-eval}
    \vspace{-0.5cm}
\end{table}

\subsubsection{Misuse fixing}

The left side of Table \ref{tab:fix-eval} shows \tool's fix rate for each model and prompt. 
Similar to detection results, for all misuses (top left part), \tool still achieves the highest fix rate using the full prompt (\llmdataDir) with \texttt{llama} (51.28\%).
We found that \texttt{llama}'s fix rate is significantly higher than those of \texttt{4o-mini} and \texttt{4o} (\textit{p}\textless{}0.05).
We observe similar trends to detection where \texttt{4o-mini} still achieves its best performance with \llmb (43.59\%).
For fixing, we find that only providing directives (\llmdir) worked best for \texttt{4o}, achieving 46.15\% fix rate across all misuses (top part of Table \ref{tab:fix-eval}). However, for the subset of \datapoints where the misused API had a directive, the full prompt performed better, reaching 63.16\% (bottom part of Table \ref{tab:fix-eval}).

\begin{findingenv}{RQ1 findings}{finding:rq1-results}
    Our results show that the majority of the models have higher detection $F_1$-score with \llmdataDir. 
    Among the three models, \texttt{llama} with \llmdataDir has both the best detection $F_1$-score (61.18\%) and the fix rate (51.28\%).
\end{findingenv}
\vspace{-0.7cm}
\section{RQ2: Does few-shot prompting improve LLMs' API misuse detection and fixing?}
\label{sec:few-shot}

\vspace{-0.2cm}
\subsection{RQ2 Setup}
\vspace{-0.2cm}
To answer RQ2, we experiment with few-shot prompting where we provide solved examples to the models.
Based on the zero-shot prompting results, we select the \llmdataDir template to proceed with few-shot prompting. 
To include in the few-shot prompt, we create two new examples (available on our artifact), one correct and one incorrect, that use APIs from \texttt{sckikit-learn} and \texttt{pandas}, respectively.
These two examples are inspired by the misuses in the dataset, but neither direct replicas nor reuse any of the misused APIs in \esemdataset.
We include these examples along with the corresponding responses at Line 4 in Figure \ref{fig:full-prompt-simpleimputer}.
We refer to this prompt as \llmfewshot.

\vspace{-0.1cm}
\subsection{RQ2 Results}
\label{sec:rq2-results}
\vspace{-0.2cm}

\subsubsection{Misuse detection}
The last column of Table \ref{tab:detect-eval} shows the results of \llmfewshot.  
If \llmfewshot achieves a higher score than the highest performing zero-shot prompt in that row, we show it in bold.
Overall, we find that misuse detection using few-shot prompting produces varied results.
It improved $F_1$-score detection for both \texttt{4o-mini} and \texttt{4o} (both statistically significant), although the increase for \texttt{4o} is much higher. 
On the other hand, \texttt{llama}'s few-shot prompt resulted in a slight drop ($<$1\%).
Overall, \texttt{llama}'s zero-shot \llmdataDir still achieved the highest detection $F_1$-score across all variations.

\subsubsection{Misuse fixing}
Looking at all misuses in Table~\ref{tab:fix-eval}, we find that few-shot examples significantly improved \texttt{4o}'s fix rate and significantly hindered \texttt{4o-mini}'s performance (\textit{p}\textless{}0.05). In contrast, the fix rate for \texttt{llama} remained effectively unchanged, with no statistically significant difference (\textit{p}=0.714). Compared to \texttt{llama}, \texttt{4o}'s fixing ability is also not significantly different (\textit{p}=0.790).

\begin{findingenv}{RQ2 findings}{finding:rq2-results}
Few-shot prompting leads to varied results across models. It significantly improves both detection and fixing performance for \texttt{4o}, but negatively affects detection for \texttt{llama} and has no impact on its fixing performance.
\end{findingenv}

\vspace{-0.3cm}
\section{RQ3: \rqthreetext}
\label{sec:dl-misuses}
\vspace{-0.3cm}

\subsection{RQ3 Setup}
\vspace{-0.1cm}
Since our approach (i.e., adding directives and data) based on the conclusions of Galappaththi et al.~\cite{dc-misuse-smr:24}, running \tool on additional libraries gives us insights into generalizability.
Accordingly, we consider \dldataset~\cite{dl-api-misuse-llm:wei2024} dataset described in Section~\ref{sec:related-work}.
To apply \tool, we require minimal reproducible examples, API directives when available, and information about variables processed by the API. Since the code in \dldataset comes from larger repositories and is not readily runnable, we select a sample of misuses to reduce the manual effort of creating minimal examples.

\paragraph{Sampling DL misuses to reproduce}
The 891 misuses in \dldataset span various misuse types and root causes, the combination of which can be divided into stratas. 
Accordingly, we follow a stratified sampling approach to diversify the selected DL misuses to cover all stratas. 
We first discard misuses of the violation type ``outdated'' and the root cause ``deprecation management errors'' to be consistent with \esemdataset as considering deprecation as a misuse is controversial~\cite{dc-misuse-smr:24}.
This leaves 395 misuses.
Based on a 95\% confidence interval and 10\% error margin, we need a sample size of 78 misuses.

Since reproducing misuses is time-consuming, we set a cap of 30 hours on our reproduction task. During this time period, we managed to review 73 misuses covering all strata. We disregard 52 misuses that were logical bugs. 
For example, Wei \etal~\cite{dl-api-misuse-llm:wei2024} considered this change--\texttt{return relu(Y+X)} to \texttt{return relu(Y)}\cite{dl-logical-bug}--as a parameter replacement. But we observe that the line before the return statement is \texttt{Y += X}. Thus, we conclude the change is a logical bug fix instead of a misuse fix. We also discarded 6 misuses whose reproduction step took more than twenty minutes, which is the maximum time we spent on a single misuse. Overall, we reproduced 15 misuses and confirmed their type and root cause. Twelve of these misuses are data dependent and 7 have API directives.  

Similar to \esemdataset, we add a correct version of each misuse, totaling 30 \datapoints. 
In this RQ, we use \texttt{llama} with the \llmdataDir prompt based on its highest $F_1$-score in RQ1. 

\subsection{RQ3 Results}
\vspace{-0.2cm}

Given the 30 DL \datapoints, we find that the detection $F_1$-score of \tool is 48\%, it also fixed 4 misuses resulting in a fix rate of 26.67\%.
This suggests that \tool works for other \dsml libraries.
However, given that \tool's detection $F_1$-score on \esemdataset was 61.18\% (Table~\ref{tab:detect-eval}), it seems that detecting DL API misuses is more challenging.
This is not surprising as Wei et al.~\cite{dl-api-misuse-llm:wei2024} report a recall of 16.49\% and a precision of 32.33\% on their complete data set.
It is interesting to note that if we focus only on the 14 code snippets where the used API had a directive, \tool's detection $F_1$-score increases to 66.67\%, closely matching its $F_1$-score of 68.18\% on \esemdataset, as shown at the bottom of Table~\ref{tab:detect-eval}.
Three of the four misuses \tool fixed were misuses with API directives, further highlighting the importance of directives in misuse detection.

To further contextualize our results, we look at \dltool's results to see how it performed on the 15 misuses we reproduced. 
We find that only four of our 15 reproduced misuses overlapped with those in Wei \etal~\cite{dl-api-misuse-llm:wei2024}'s test set. 
Two of these four common misuses were not detected by either tool, one was detected and fixed only by \tool, while one was detected and fixed by only \dltool.
One of the undetected misuses involved a missing \texttt{NaN} check on the output of the neural network's loss function, caused by an algorithmic error.
The other undetected misuse was the absence of a \texttt{to()} call to set a tensor's device to match that of another interacting tensor. 
The misuse that only \tool detected and fixed involved passing a list to tensors when calculating the mean along the rows. 
Since \texttt{mean()} requires a tensor as input, the tensors in the list needed to be stacked first using \texttt{stack()}. 
The misuse that only \dltool detected and fixed involved a project specific type check.

\begin{findingenv}{RQ3 findings}{finding:rq3-results}
    \tool's detection $F_1$-score on the DL sample dataset is 48\%, with a 26.67\% fix rate. 
    It detected and fixed a DL misuse that \dltool did not detect.
\end{findingenv}
\vspace{-0.5cm}
\section{RQ4: \rqfourtext}
\label{sec:llm-api-det}

\subsection{RQ4 Setup}
\vspace{-0.2cm}
We also compare how \dltool~\cite{dl-api-misuse-llm:wei2024}, described in Section~\ref{sec:related-work}, performs on \esemdataset.
As a reference point, recall that as per their reported results on deep learning misuses, \dltool detected 48 misuses (16.49\% recall rate) and fixed 10/48 (22.83\%) of them~\cite{dl-api-misuse-llm:wei2024}. As a proportion of the total number of evaluated misuses, its fix rate is $\sim$3\% (10/291).

To use \dltool on \esemdataset, we need to use some misuses for the rule generation phase that are different from the misuses we use for evaluation.
Since Wei et al. had multiple instances of the same/similar misuses in \dldataset, they created a single split between rule generation data and evaluation data.
However, \esemdataset mainly contains \textit{unique} misuses.
To still enable the comparison, we come up with two setups.
In the first setup, we apply \dltool{} to our data using the rules that Wei et al.~\cite{dl-api-misuse-llm:wei2024} created from PyTorch and TensorFlow misuses. 
Such an experiment can reveal if the misuse rules they collected are generalizable to other \ds libraries.
In the second setup, we create API misuse rules from the \esemdataset dataset using a 5-fold cross-validation and report average recall for misuse detection and average fix rate for fixing misuses.

We use \texttt{4o-mini} instead of the now-deprecated \texttt{gpt-3.5} used by Wei \etal~\cite{dl-api-misuse-llm:wei2024}, as it is the closest available alternative. Reproducing \dltool with \texttt{4o-mini} yields a recall of 17.81\%, closely matching the originally reported 16.49\%. This result validates our reproduction setup.
Similar to Wei \etal~\cite{dl-api-misuse-llm:wei2024}, we only report recall for detection by applying \dltool to the 38 misuses in the \esemdataset.

\subsection{RQ4 Results}
\vspace{-0.2cm}

In the first setup, we find that \dltool could not detect any of the 38 misuses. 
Even though the \esemdataset contains misuses caused by similar root causes, such as data conversion errors due to a missing \texttt{dtype} argument in NumPy APIs (which is also a common misuse pattern in DL APIs), \dltool failed to detect them. 
We observed that the rules generated by \dltool contain specific DL API names and when the tool tries to find semantically matching rules to apply for a given code snippet, it first applies a filtering step where it looks for overlapping API names in the code snippet and rules. 
As a result, \dltool fails to find matching rules for code snippets in \esemdataset, since none use DL APIs.

In the second setup with 5-fold cross validation, \dltool detects only four unique misuses across all folds, achieving 11.19\% average recall.
We note that all these four misuses were of Matplotlib APIs.
As shown in the results of RQ1 in Table~\ref{tab:detect-eval}, \llmdataDir with \texttt{4o-mini} achieves a 46.15\% recall, representing a 34.96\% increase over \dltool.  

\dltool did not fix any of the four detected misuses. Upon further inspection of the candidate fix rules selected by the LLM, we find that while the rules were related to Matplotlib, none were applicable--possibly due to the unique nature of each misuse in \esemdataset. This highlights a key advantage of \tool: it does not rely on prior examples of API misuse fixes.

\begin{findingenv}{RQ4 findings}{finding:rq4-results}
\tool detects $\sim$35\% more misuses than \dltool on \esemdataset.
Additionally, \dltool was not able to fix any of the misuses it detected.
\end{findingenv}
\vspace{-0.7cm}
\section{RQ5: Can \toolagentic detect and fix API misuses?}
\label{sec:toolcalling}
\vspace{-0.3cm}

\subsection{RQ5 Setup}
\vspace{-0.2cm}
All the above results suggest that \tool is effective at detecting and fixing misuses.
However, our setup so far has been idealistic where we know which API to provide information about, and we have this information readily available.
To investigate a more realistic setup, we implement \toolagentic which relies solely on function calling (without agentic frameworks such as ReAct~\cite{react:23}) to allow the LLM to request API documentation and variable information on demand. 
We use \llmb that contains only the code snippet and the library used in it and  modify the system message to inform the LLM that it can request information by calling functions.
We implement two functions that allow the LLM to request additional information, without restricting the number of times they can be called.

\subsubsection{Function 1: \texttt{get\_variable\_info}}
This function takes a variable name and line number as input. Based on the input parameters, it then analyzes the abstract syntax tree (AST) of the current code snippet to determine where to insert instrumentation.
The inserted code uses appropriate print statements similar to those manually added in Section~\ref{sec:data-prep} to extract relevant information.
The function then executes the instrumented code and returns the variable’s type, sample values, and any additional available details.

\subsubsection{Function 2: \texttt{get\_api\_documentation}}
This function takes an API name as input and returns its full documentation by querying a locally maintained index covering five libraries in \esemdataset.
Note that this function returns the full documentation of a given API, rather than a specific API directive.
Since we do not know which directive is relevant and automatically extracting directives from documentation is challenging~\cite{Directives-Monperrus:2012, Rec-API-Robillard:2015, API-doc-patterns-Maalej:2013}, we provide full API documentation and leave the LLM to decide how to use that information.

Since \texttt{4o} and \texttt{llama} had no statistically significant differences in RQ1, we use both models. In contrast to \llmdataDir in RQ1 where the prompt includes exact pre-determined information catered to each misuse, RQ5 setup lets the LLM decide which information it needs and for which exact variable or API.

\vspace{-0.1cm}
\subsection{RQ5 Results}
\vspace{-0.2cm}

\subsubsection{Misuse detection}

We find that \toolagentic achieves an $F_1$-score of 43.33\% with \texttt{llama} and 48.65\% with \texttt{4o} for misuse detection, with the latter being significantly higher (\textit{p}\textless{}0.05).
These results suggest that, overall, the LLM (\texttt{4o} in this case) can effectively request the information it needs to determine API misuse. This shows that our approach could be integrated into a practical setting without pre-defined knowledge of the misused API.

On the other hand, \toolagentic achieves a $\sim$7\% significantly lower $F_1$-score than \tool with \texttt{4o} on \llmdataDir (\textit{p}\textless{}0.05).
To understand this drop, we examine the function calls made by \toolagentic. 
In total, \texttt{4o} made 136 function calls, averaging 1-2 calls per \datapoint. 
Each \datapoint in our dataset contains 1-23 APIs (median 6), belonging to the targeted library and other libraries.
\texttt{4o} called for documentation 111 times, 26 of which were made for the corresponding potentially misused API.
This shows that the LLM often requested the relevant API information. 
However, the decreased $F_1$-score suggests that providing the full API documentation is less effective than directly offering the relevant directive.
\texttt{4o} called \texttt{get\_variable\_info} 25 times, 14 of which were for the variable processed by potentially misused API.
In comparison to \texttt{4o}, \texttt{llama} also made 111 function calls, all of which were for API documentation. Only 15 of 111 were for the potentially misused API which possibly explains its lower $F_1$-score relative to \texttt{4o}. 

\subsubsection{Misuse fixing}
In misuse fixing, \texttt{4o} and \texttt{llama} achieved fix rates of 39.47\% and 21.05\%, respectively. Compared to the best zero-shot fix rate in Table~\ref{tab:fix-eval}, the 12\% drop for \texttt{4o} is not statistically significant (\textit{p}=0.941). In contrast, \texttt{llama}'s 30\% decrease relative to its best zero-shot performance is statistically significant (\textit{p}\textless{}0.05). 

\begin{findingenv}{RQ5 findings}{finding:rq5-results}
\toolagentic  achieved a detection $F_1$-score of 48.65\% and a fix rate of 39.47\% with \texttt{4o}.
Out of 136 function calls, 40 were for potentially useful information.
\end{findingenv}

\vspace{-0.3cm}
\section{Discussion}
\label{sec:discussion}
\vspace{-0.2cm}

In this section, we discuss the specific implications of our work for these developers and their support tools, as well as for researchers and library authors.

\paragraph{Implications for tooling and developers} 
Our findings demonstrate the strong potential of using LLMs for detecting and fixing API misuses with a \dsml nature. 
\tool eliminates the need for prior knowledge of fixing misuses, which are requirements that have traditionally constrained both conventional API misuse detection tools~\cite{Misuse-Monperrus:2013, PR-Miner-Li:2005, Colibri-ML-Lindig:2015, GrouMiner-Nguyen:2009, Jaddet-Wasylkowski:2007, MuDetect-Amann:2019} and even more recent machine-learning/LLM-based solutions~\cite{llm-crypto:24, dl-api-misuse-llm:wei2024}. 
Our experiments show that, when provided with the right information, such as API directives and dynamic data information, LLMs can provide effective tooling off-the-shelf for this task.
While the performance of \toolagentic suggests practical promise for developer use, the decreased detection and fixing scores indicate that further improvements by researchers may be needed, which we discuss next.

\paragraph{Implications for researchers and library authors}
To the best of our knowledge, this paper is the first to explore LLMs' agentic behavior in the context of API misuse.
Our findings raise several challenges for future research.

First, while the LLM often requested API documentation, we noticed hallucinated requests for APIs that either did not exist or were not used in the current code snippet. 
Researchers can explore potential validation steps before executing the called functions or feedback loops that inform the LLM of its meaningless requests. 

Second, even in cases where the LLM requested documentation for the correct API, it often struggled to effectively utilize this information for accurate detection and repair.
This suggests that current API documentation formats, primarily designed for human consumption, pose a challenge for LLMs. 
We also observed cases where the critical API directive was not in the retrieved API reference documentation page but in other tutorial or example pages in the library's documentation.
While automatically identifying API directives remains an open and difficult problem~\cite{Directives-Monperrus:2012, Rec-API-Robillard:2015, API-doc-patterns-Maalej:2013,API-KG-Ren:2020, API-graph-Zeng:2021}, our findings underscore the need for further work in this area.
With developers' increased reliance on LLMs, library authors could also focus on augmenting API documentation with LLM-friendly structured formats to improve machine readability. 

\vspace{-0.2cm}
\section{Threats to Validity}
\vspace{-0.2cm}

\textit{Internal Validity.} 
We manually assess the correctness of the LLM's fixes and explanations.
While manual processes are often subjective, the dataset contains detailed descriptions of the problem and the error messages or incorrect output reflecting the problem.
Thus, we could precisely assess if the fix addressed the problem, especially with re-running the code.

\textit{Construct Validity.} 
A model may label code as incorrect but suggest an unrelated fix, which would inflate detection rates without reflecting true understanding.
To reduce this risk, we require that the explanation aligns with the misuse description in the dataset to consider it a successful detection. Additionally, since a misuse can have multiple valid fixes, we run the patched code to verify that the misuse is resolved, regardless of the implementation.

It is possible that our dataset might overlap with LLM's training data. To reduce this risk, we did not directly reuse code from \so posts or \gh commits when constructing the code snippets. This makes it less likely that models succeed merely by memorizing training data rather than demonstrating actual misuse detection ability.

\textit{Conclusion Validity.}
Our dataset size may limit the reliability of the comparisons. Additionally, using a single measurement to compare treatments, such as LLM variants and prompts, may provide a misleading picture. To reduce this risk, we compare distributions of measurements through bootstrapping, enabling a statistically grounded analysis.

\textit{External Validity.} 
We evaluate only three LLMs in this work.
Experimenting with additional LLMs may produce different results.
However, we tried to select representative LLMs that have been shown to perform well in previous work \cite{llm-static-bug-detect:24, llm-context-learning:24}.
We mix between different size proprietary LLMs from OpenAI as well as an open-source model with a large size.

LLMs are inherently non-deterministic, meaning their outputs can vary across different runs. This variability can be particularly challenging in our evaluation process, which involved two manual steps: assessing free-text explanations and executing generated code. 
Due to the resource-intensive nature of these evaluations, instead of running multiple times for each data point, we set the temperature to 0.0, to produce more focused and deterministic responses as much as possible~\cite{openai-requests}. 

Our primary evaluation was conducted on the \esemdataset, a dataset with a limited number of misuses but which has the crucial advantage of providing minimally reproducible examples for thorough validation. To explore broader applicability, we extended our evaluation to a sample of DL misuses.
\vspace{-0.1cm}
\section{Conclusion}
\vspace{-0.2cm}

This study investigated LLM effectiveness in detecting and fixing API misuses in \ds libraries, which are inherently harder to detect given their \dsml nature. 
We show that providing the LLM with API directives and dynamic variable information enhances performance: the best model, \texttt{llama}, achieved a detection $F_1$-score of 61.18\% and correctly fixed 51.28\% of misuses with this added context. 
Our experiments further demonstrate that \tool is applicable beyond the initial \ds libraries, successfully detecting API misuses in deep learning libraries, even identifying cases missed by an existing LLM-based tool specifically designed for \dl misuse detection. Notably, that tool struggled when it lacked examples to learn misuse-fixing patterns.
We also explored equipping the LLM with agentic behavior for real-world settings.
While the detection $F_1$-score dropped to 48.65\% and fix score to 39.47\%, the results show promise for practical settings where no prior information about the potential misuse is known.
We also identified future directions to further pursue and improve LLM agentic behavior. 

\vspace{-0.1cm}
\section{Acknowledgement}
\vspace{-0.2cm}
We used ChatGPT~\cite{openai-chatGPT} for proofreading and content reduction, per IEEE's submission policy \cite{ieee-sumission-policy}. 
\vspace{-0.1cm}
\bibliographystyle{IEEEtran}
\bibliography{references}

\end{document}